\documentclass[runningheads]{llncs}

\usepackage{booktabs}
\usepackage{cite}
\usepackage{float}
\usepackage{url}

\usepackage{amsmath} 
\usepackage{mathtools}
\usepackage{amssymb}
\usepackage{accents}

\usepackage[flushleft]{threeparttable} 
\usepackage{adjustbox}
\usepackage{geometry}
\usepackage{xcolor}

\begin{document}

\title{Speeding up the solution of the Site and Power Assignment Problem in Wireless Networks}

\titlerunning{Site and Power Assignment in Wireless Networks}

\author{Pasquale Avella\inst{1} \and Alice Calamita\inst{2}\thanks{Corresponding author. This author has been partially supported by Fondazione Ugo Bordoni, Rome, Italy} \and Laura Palagi\inst{2}}

\authorrunning{P. Avella et al.}

\institute{DING, Università del Sannio, Piazza Roma 21, 82100, Benevento, Italy\\
\email{avella@unisannio.it}
\and
DIAG, Sapienza University of Rome, Via Ariosto 25, 00185, Rome, Italy\\
\email{\{alice.calamita,laura.palagi\}@uniroma1.it}}

\maketitle

\begin{abstract}
This paper addresses the optimal design of wireless networks through the site and power assignment problem. Given a set of candidate transmitters, this problem involves choosing optimal transmitter locations and powers to provide service coverage over a target area. In the modern context of increasing traffic, establishing suitable locations and power emissions for the transmitters in wireless networks is a relevant and challenging task due to heavy radio spectrum congestion. Traditional network design formulations are very ill-conditioned and suffer from numerical inaccuracies and limited applicability to large-scale practical scenarios. Our contribution consists of speeding up the solution of the problem under consideration by addressing its drawbacks from a modeling point of view. We propose valid cutting planes and various presolve operations to reduce the problem size and strengthen existing formulations, along with a reduction scheme based on reduced cost fixing to reduce the sources of numerical inaccuracies. Our proposals prove effective, allowing us to achieve optimality on large-scale instances obtained from a real 4G LTE network in solution times aligning well with planning windows.

\end{abstract}

\section{Introduction}

A wireless network is a telecommunications network that uses radio waves, or other wireless communication technologies, to transmit and receive data without the need for physical cables, allowing for the wireless connectivity of devices. From a design point of view, the basic elements of wireless networks are transmitters and receivers. Hence, wireless network design (WND) consists of identifying the proper locations for the transmitters and setting their operational parameters -- like frequency and/or power emission -- in such a way as to cover with service the receivers located in the area of interest.

Even if wireless networks rely on different technologies
and standards based on the service they are meant to provide, they all share a common feature: the need to reach users scattered over a vast area with a radio signal that must be strong enough to prevail against other unwanted interfering signals. The quality of service (and hence the coverage) indeed depends on the interplay of numerous signals, wanted and unwanted, generated from a large number of transmitting devices. 
The increasing traffic and the densification of the base stations along the territory have led to an increase in interfering signals. Consequently, establishing suitable locations and power emissions for all the transmitters, coexisting within a heavily congested radio spectrum, has become a challenging and relevant task. 
Indeed, in the current era of pervasive connectivity, the design of wireless networks plays a pivotal role in shaping modern societal infrastructure. 
The importance of studying wireless network design lies in its profound implications for enhancing communication and enabling technological advancements, and its influence on the evolving dynamics of global connectivity in daily lives. 

This paper addresses the design of wireless networks for 4G LTE technology and considers the frequency as fixed, thus tackling a WND problem known as the site and power assignment problem. 
This research was carried out in collaboration with the Fondazione Ugo Bordoni (FUB) \cite{FUB}, a higher education and research institution under the supervision of the Italian Ministry of Enterprises and Made in Italy (MISE) that operates in the telecommunication field, providing innovative services for government bodies. 

\paragraph{Literature overview} As wireless networks are becoming denser, due to technological advancements and increased traffic \cite{israr2021renewable}, practitioners' traditional design approach based on trial-and-error supported by simulation has exhibited many limitations. The inefficiency of this approach led to the need for optimization approaches, which resulted in been critical for lowering costs and meeting user-demanded service quality standards (see e.g., \cite{dehghan2005new, d2013gub}).
Many optimization models for WND have been investigated over the years. However, the natural formulation on which most models are based presents severe limitations since it involves numerical problems in the problem-solving phase, which emerge even in small instances. 
Indeed, the constraint matrices of these models contain coefficients that range in a huge interval, as well as large big-$M$ leading to weak bounds. In this paper we review the exact approaches proposed for WND problems, pointing out the contributions that highlight these numerical issues. We recommend \cite{avella2023compact} for a complete literature review on WND (also including heuristic approaches) and \cite{kennington2010wireless} for a thorough overview of the optimization challenges in modern WND. 

The exact approaches proposed in the literature are mainly oriented towards
non-compact formulations and row generation methods. In \cite{naoum2010nested}, a  mixed-integer linear programming (MILP) formulation is introduced, and the proposed exact solution method combines combinatorial and classical Benders decomposition and valid cuts.
In \cite{ageyev2015lte,ageyev2014optimization,bondarenko2019optimization,dmytro2015multi}, mixed-integer formulations are used to solve randomly generated instances through standard MIP solvers. 
In \cite{d2013gub}, a non-compact 0-1 formulation is investigated. The solution algorithm is based on a row-generation method. The same authors of \cite{d2013gub} present a 0-1 model for a WND variant linked to the feasible server assignment problem in \cite{d2011negative}. In \cite{capone2011new}, a non-compact formulation is proposed for the maximum link activation problem; the formulation uses cover inequalities to replace
the source of numerical instability. In \cite{d2016towards}, the source of numerical issues in WND is deeply investigated, and the use of numerically safe LP solvers is suggested to make the solutions reliable.
In \cite{avella2023compact}, a compact reformulation for the siting problem has been proposed that allows for the exact solution of large instances. The papers explicitly addressing numerical issues are \cite{d2013gub,d2011negative,capone2011new,d2016towards,avella2023compact}.

\paragraph{Main motivation and contributions}
Traditional solution methods, employing (mixed-)integer linear programs with (very) ill-conditioned coefficient matrices, suffer from numerical inaccuracies and limited applicability to large-scale practical scenarios. Indeed, the traditional modeling choice typically includes big-$M$ coefficients to model coverage conditions, leading to very weak linear relaxations and solutions -- returned by state-of-the-art MIP solvers -- typically far from the optimum \cite{d2012pure}. 
Our contribution consists of speeding up the solution of the problem under consideration, by addressing its drawbacks, from a modeling point of view. 
Specifically, we discuss how to improve the natural formulation of the WND problem proposed in the literature by \begin{enumerate}
    \item introducing several presolve operations to reduce the number of problem variables and overall problem size; 
    \item providing valid cliques and variable upper bounds to accelerate solution times;
    \item proposing an aggressive reduction scheme based on a reduced cost fixing procedure that strengthens the formulation by reducing the big-$M$ values.
\end{enumerate} 
Although reduced cost fixing is a well-known technique, its application to this problem has never been investigated before (as far as we know) and shows significant potential thanks to the positive impact on the reduction of numerical problems. To improve this technique, we provide a heuristic, producing near-optimal solutions very quickly. Our proposals to reduce memory and numerical issues will allow a rapid solution of large WND instances in line with the times required in the planning phase.

The remainder of this paper is organized as follows. Section \ref{ch_formulation} presents the problem statement and its mathematical formulation. In Section \ref{ch_contribution}, our contribution to accelerating the problem solution is given. In Section \ref{ch_results}, computational results are discussed. Conclusions are provided in Section \ref{ch_conclusions}.

\section{Problem Formulation}
\label{ch_formulation}

A wireless network consists of radio transmitters distributing service (i.e., wireless connection) to a target area. 
The target area is usually partitioned into elementary areas, called testpoints, in line with the recommendations of the telecommunications regulatory bodies. Each testpoint is considered a representative receiver of all the users in the elementary area.
Testpoints receive signals from all the transmitters. The power received is classified as serving power if it relates to the signal emitted by the transmitter serving the testpoint; otherwise, it is classified as interfering power (see 4G LTE standard \cite{rumney2013lte}). A testpoint is regarded as served (or covered) by a base station if the ratio of the serving power to the sum of the interfering powers and noise power
(Signal-to-Interference-plus-Noise Ratio or SINR) is above a threshold \cite{rappaport1996wireless}, whose value
depends on  the desired quality of
service.

We assume the frequency channel is given and equal for all the transmitters. Having a fixed frequency is a straightforward assumption: for different frequency channels the problem decomposes as there is no interference among non-co-channel signals. We also assume that the power emissions of the activated transmitters can be represented by a finite set of power values, which fits with the standard network planning practice of considering a small number of discrete power values. This practice of power discretization for modeling purposes has been introduced in \cite{d2013gub}. 

Let $\mathcal{B}$ be the finite set of potential transmitters and $\mathcal{T}$ be the finite set of receivers located at the testpoints. Let $\mathcal P = \{P_1, \dots,P_{|\mathcal P|}\}$ be the finite set of feasible power values assumed by the activated transmitters, with $P_1>0$ and $P_{|\mathcal P|} = P_{max}$. Hence $\mathcal L = \{1, \dots, |\mathcal P|\}$ is the finite set of power value indices (or simply power levels).
We introduce the variables 
\begin{equation*}
    z_{bl}=\left\{\begin{array}{ll}
1     & \mbox{if transmitter $b$ is emitting at power $P_l$} \\
0     &  \mbox{otherwise}
\end{array}\right.\hfill \qquad b\in\mathcal{B},\, l\in\mathcal{L}
\end{equation*}
and
\begin{equation*}
x_{tb}=\left\{\begin{array}{ll}
1     & \mbox{if testpoint $t$ is served by transmitter $b$} \\
0     &  \mbox{otherwise.}
\end{array}\right.\hfill \qquad  b\in\mathcal{B},\, t\in\mathcal{T}
\end{equation*}

To enforce the choice of only one (strictly positive) power level for each activated transmitter we use
\begin{equation}
\sum_{l \in  \mathcal{L}} z_{bl} \leq 1 \qquad b \in  \mathcal{B}. \label{constr_power}
\end{equation}

The mathematical formulation of the WND problem contains the so-called SINR inequalities used to assess service coverage conditions. We recall that a testpoint $t  \in \mathcal T$ is regarded as covered by a base station $\beta \in \mathcal B$ if the SINR$_{t\beta}$ is above a threshold. The SINR$_{t\beta}$ is given by the ratio of the serving power coming from $\beta$ and the sum of the interfering powers and noise power measured in $t$. The serving power is given by the product of the power emitted by the serving base station $\beta$ and the fading coefficient, modeling the reduction of power in the signal from $\beta$ to $t$. The interfering power is given by the sum of all interfering contributions measured at the testpoint $t$. These interfering contributions are modeled as the serving power, with the unique difference that the signal is emitted by all the activated base stations, except for the serving one (i.e., except for $\beta$).
To formulate the SINR inequalities, we refer to the discrete big-$M$ formulation reported in \cite{d2013gub}, which considers a discretization of the power range. 
Let $\tilde a_{tb} > 0$ be the fading coefficient applied to the signal received in $t \in \mathcal T$
and emitted by $b \in \mathcal B$. Let $\mu > 0$ be the system noise. Then a receiver $t$ is served by a base station $\beta \in\ \mathcal{B}$ if the SINR$_{t\beta}$ is above a given SINR threshold $\delta > 0$, namely
\begin{equation} 
\label{SINR}
\frac{\tilde {a}_{t\beta}\displaystyle\sum_{l \in  \mathcal{L}} P_l z_{\beta l}}{\mu + \displaystyle\sum_{b \in  \mathcal{B} \setminus \{\beta\}} \tilde a_{tb} \sum_{l \in  \mathcal{L}} P_l z_{bl}} \geq \delta \qquad t \in \mathcal T, \beta \in \mathcal B:\ x_{t \beta} = 1 
\end{equation}
where the numerator represents the serving signal in $t$ (coming from $\beta$), and the denominator is the sum of the noise and the interfering signals in $t$ (coming from all $b \neq \beta$).
Following \cite{d2013gub}, we can rewrite the SINR condition \eqref{SINR} through the big-$M$ constraints 
\begin{equation} 
\tilde a_{t\beta} \sum_{l \in \mathcal L} P_l z_{\beta l} - \delta \sum_{b \in  \mathcal{B} \setminus \{\beta\}} \tilde {a}_{tb} \sum_{l \in \mathcal L} P_l z_{bl} \geq \delta \mu - M_{t \beta} (1 - x_{t \beta}) \qquad t \in  \mathcal{T}, \beta \in  \mathcal{B}  
\label{sinr_ineq_general}
\end{equation}
where $M_{t\beta}$ is a large (strictly) positive constant.
When $x_{t \beta}=1$, \eqref{sinr_ineq_general} reduces to \eqref{SINR}; when 
$x_{t \beta}=0$ and $M_{t\beta}$ is sufficiently large,
 \eqref{sinr_ineq_general}
  becomes redundant. We can set, e.g., 
  \begin{equation}\label{bigM}
  M_{t \beta} = \delta \mu + \delta P_{max} \sum_{b \in  \mathcal{B} \setminus \{\beta\}} \tilde a_{tb}.
  \end{equation}
Note that we can claim that a testpoint $t \in \mathcal T$ is covered if and only if there exists at least one $(t, \beta)$ with $\beta \in \mathcal B$ that can satisfy \eqref{sinr_ineq_general} with $x_{t \beta} = 1$.

A constraint to express a minimum service coverage of the territory is included.
We assume that each testpoint weights $r_t \in \mathbb R$ to account for the fact that testpoints can represent a different number of users or be more or less crucial in service coverage.
A minimum coverage level $r\in \mathbb R$ of the testpoint is enforced by the constraint:
\begin{equation}
\sum_{t \in  \mathcal{T}} r_t \sum_{b \in  \mathcal{B}} x_{tb} \geq r. \label{constr_coverage}
\end{equation}
When $r_t$ are all equal to 1, $r$ represents the minimum number of testpoint to be covered, and the constraint above corresponds to a territorial coverage.
When $r_t \in [0,1]$ can be interpreted as the fractions of the population living in the elementary area $t \in \mathcal T$, then $r\in [0,1]$ represents the fraction of the population to be covered by the service and the constraint represents a population coverage.

Each testpoint must be covered by at most one serving base station, namely
\begin{equation}
 \sum_{b \in  \mathcal{B}} x_{tb} \leq 1 \qquad  t \in  \mathcal{T} \label{constr_receivers}.
\end{equation}

The objective functions proposed in the literature of WND are several, going from the maximization of the coverage to the maximization of the quality of service. According to the FUB, a considerable goal to pursue nowadays is the citizens' welfare; therefore, the model we refer to aims at identifying solutions with low environmental impact in terms of electromagnetic pollution and/or power consumption. Reducing electromagnetic pollution indeed involves reducing the power emitted by the transmitters \cite{chiaraviglio2018planning}. Hence, we aim to minimize the total number of activated base stations with a penalization on the use of stronger power levels: the cost associated with the use of a power level equal to $l \in \mathcal L$, namely $c_l$, is greater the greater is $P_l$.

Thus, a natural formulation of the WND is the following 0-1 LP model:
\begin{equation}\label{eq:general}
    \begin{array}{rlr}
\min_{x,z} \hspace{2mm}& \displaystyle\sum_{b \in  \mathcal{B}}\sum_{l \in  \mathcal{L}} c_{l} z_{bl}
\\
&(x,z)\in S
\end{array}
\end{equation}
where the feasible region $S$ is defined as $$S=\left\{(x,z)\in\{0,1\}^{n+m}: \ \mbox{satisfying }  \eqref{constr_power},\eqref{sinr_ineq_general}, \eqref{constr_coverage},  \eqref{constr_receivers}\right\} $$
with $x=(x_{t b})_{t \in  \mathcal{T},\,b \in  \mathcal{B} }, z=(z_{bl})_{b \in  \mathcal{B}, \,l \in \mathcal L } $ and $n=|\mathcal{T}| \times |\mathcal{B}|, m=|\mathcal{B}| \times |\mathcal{L}|$.

Sets, parameters and variables used in the model are summarized in Table \ref{table:par_var_set}.
\begin{table} [h] 
\caption{Sets, parameters and variables}
\begin{tabular}{ll}
  \textbf{Symbol} & \textbf{Notation} \\
  \hline
    $\mathcal{B}$ & Set of potential transmitters\\
    $\mathcal{T}$ & Set of receivers\\
    $\mathcal{P}$ & Set of feasible power values\\
    $\mathcal{L}$ & Set of feasible power indices\\
    \hline
    $\tilde {a}_{tb}$ & Fading coefficient applied to the signal from transmitter $b\in \mathcal B$ to testpoint $t \in \mathcal T$\\
    $c_l$ & Cost of using power level $l \in \mathcal L$\\
    $r_t$ & Weight of testpoint $t \in \mathcal T$\\
    $r$ & Coverage level\\
    $\delta$ & SINR threshold \\ 
    $\mu$ & System noise\\
    \hline
    $z_{bl}$ & 0-1 Variable representing if transmitter $b \in \mathcal B$ is emitting at power level $l \in \mathcal L$\\
    ${x}_{tb}$ & 0-1 Variable representing if testpoint $t \in \mathcal T$ is served by transmitter $b \in \mathcal B$\\
  \hline
\end{tabular}
\label{table:par_var_set}
\end{table}

\medskip

In principle, MIP solvers can solve model \eqref{eq:general}. However, it is well-known that the following issues may arise, as widely described, e.g., in \cite{d2013gub,d2016towards}:
\begin{itemize}
 \item[\textbullet] the power received in each testpoint lies in a large interval, from very small values ($10^{-7}$) to huge ($10^5$), which makes the range of the coefficients $\tilde a_{tb}$ in the constraint matrix very large and the solution process numerically unstable and possibly affected by error; 
    \item[\textbullet] the big-$M$ coefficients lead to poor quality bounds that impact the effectiveness of standard solution procedures;
    \item[\textbullet] real-world problems lead to models with a large number of variables and constraints.
\end{itemize}
These issues make the solution of real-life instances of this problem very challenging. Therefore, our study aims at making the solution of this problem more efficient and fast, hence practicable.

\section{Our Contribution}
\label{ch_contribution}

We mentioned that practical WND problems are hard to solve using optimal procedures as numerical and memory issues arise even in small instances of the problem.
In this section, we discuss how to strengthen natural formulations of the problem by means of presolve operations, valid cutting planes, and a coefficient tightening procedure. All these operations will speed up the solution of the problem and reduce its size, as shown in the computational results section.

\subsection{Presolve operations}\label{ch_presolve_variable}
Reducing the model size is crucial as real-life instances typically involve many variables and constraints. In this section we describe several operations that allow us to reduce the size of the problem by eliminating some $x_{tb}$ and $z_{bl}$ variables a priori. Note that the elimination of the $x_{tb}$ variables also leads to the a priori elimination of the corresponding SINR$_{tb}$ constraint \eqref{sinr_ineq_general}.

\subsubsection{Reducing the number of servers}
Usually, for reasons related to signal quality, only a certain number of transmitters (the ones emitting the strongest signals received from the testpoint) are generally considered as possible servers for a given testpoint.
Therefore, we establish a number of possible servers for each testpoint a priori. Namely, for each testpoint $t$, we select a subset of servers $S_t$, corresponding to the transmitters emitting the strongest signals received in $t$.

Then, to further reduce the size of the problem, we delete all the variables modeling transmitter-receiver pairs which any feasible
solution would exclude. Specifically, suppose we exclude the possibility of serving the full target area with a single transmitter, indeed this case is trivial and would not require the use of optimization to be solved. In that case, the best-case scenario is the one with only two transmitters deployed, where one transmitter works as a server and the other as an interferer.

Hence, for each testpoint $t$, we fix $x_{ts} = 0$ for all those transmitters $s \in S_t$ such that the SINR$_{ts}$ is below the threshold $\delta$ for each possible single interferer $b \in \mathcal B: b \neq s$. Namely, we eliminate all $x_{ts}$ such that 
$$\text{SINR}_{ts} = \frac{\tilde a_{ts}P_i}{\mu + \tilde a_{tb}P_j} < \delta \qquad \forall b \in \mathcal B \setminus \{s\}, \forall P_i, P_j \in \mathcal P$$
which can be easily verified directly with
$$\text{SINR}^{max}_{ts} = \frac{\tilde a_{ts}P_{max}}{\mu + \tilde a_{th}P_{min}} < \delta \qquad \text{with} \quad \tilde a_{th} \coloneqq \min_{b \in \mathcal B \setminus \{s\}}{\{\tilde a_{tb}\}}.$$

\subsubsection{Reducing power levels}

To reduce the size of the problem, we also select a subset of transmitter power levels which any feasible solution would exclude. Indeed, excluding again the possibility of serving the full target area with a single transmitter, we can consider the best-case scenario as the one with only two transmitters deployed, one server and one interferer. Hence, for each transmitter $b \in \mathcal B$ and level of power $l \in \mathcal L$, we fix $z_{bl} = 0$ for all those $(b,l)$ such that the SINR is below the threshold $\delta$ for each possible interferer $\beta \in \mathcal B: \beta \neq b$ and for each testpoint $t \in \mathcal T$. Namely, we eliminate all $z_{bl}$ such that 
$$\text{SINR}_{tb} = \frac{\tilde a_{tb}P_l}{\mu + \tilde a_{t \beta}P_j} < \delta \qquad \forall t \in \mathcal T, \forall \beta \in \mathcal B \setminus \{b\}, \forall P_j \in \mathcal P$$
which can be easily verified directly with
$$\text{SINR}^{max}_{tb} = \frac{\tilde a_{tb}P_l}{\mu + \tilde a_{th}P_{min}} < \delta \qquad \forall t \in \mathcal T,\, \text{with} \quad \tilde a_{th} \coloneqq \min_{\beta\in \mathcal B \setminus \{b\}}{\{\tilde a_{t \beta}\}}.$$

\subsubsection{Heuristic sparsification}
We perform a heuristic sparsification to deal with the numerical issues arising from the coefficients of the SINR
inequalities. Specifically, we set a minimum threshold $\varepsilon$ on the received power below which the received power can be considered null. Namely, for each $t \in \mathcal T,\, b \in \mathcal B,\, P_i \in \mathcal P$ we set
$$
\tilde a_{tb}P_i = \begin{cases}
\tilde a_{tb}P_i &  \text{if } \tilde a_{tb}P_i \geq \varepsilon\\
0 & \text{otherwise.}
\end{cases}$$
This allows us to reduce the size of the problem by eliminating some $x_{tb}$ variables a priori.

\subsection{Cutting planes}\label{ch_cutting_planes_variable}
A standard procedure for solving 0-1 LPs is the branch-and-bound algorithm, which can significantly be improved by cutting planes, i.e., inequalities that are valid for all integer solutions but not for
some solutions of the linear relaxation. By means of such inequalities, fractional linear relaxation solutions can be cut off.
Valid inequalities are internally generated by state-of-the-art MIP solvers. However, MIP solvers cannot take advantage of the particular problem structure known to the user. For the problem at hand, we identify some problem-specific cutting planes, including variable upper bounds and families of clique inequalities, that we provide in this section.

\subsubsection{Families of clique inequalities}
Suppose again to exclude the possibility of serving the full target area with a single transmitter and consider the best-case scenario as the one with only two transmitters deployed, one server and one interferer.
We observe that we can exclude (i) potential levels of power for a certain transmitter $b \in \mathcal B$ and (ii) potential serving signals for a certain testpoint $t \in \mathcal T$
simply considering the minimum SINR required in $t$.

Hence, for each testpoint $t \in \mathcal T$, interferer $b \in \mathcal B$ with power $P_l \in \mathcal 
P$ and server $\beta \in \mathcal B$ such that the SINR measured in $t$ and that considers as the only interferer $b$ is always below the threshold $\delta$, i.e. 
\begin{equation*}
\frac{\tilde a_{t \beta}P_{i}}{\mu + \tilde a_{tb}P_l} < \delta \quad \forall P_i \in \mathcal P
\end{equation*}
which can be easily verified directly with
\begin{equation}
\label{condition_clique_general}
    \frac{\tilde a_{t \beta}P_{max}}{\mu + \tilde a_{tb}P_l} < \delta
\end{equation}
we can exclude the possibility that $b$ is activated at a power level equal to $l$ and simultaneously $\beta$ serves $t$ using
\begin{equation}
\label{clique_semplice_generale}
    z_{bl} + x_{t \beta} \leq 1.
\end{equation}
Using \eqref{constr_power} and \eqref{constr_receivers}, we can strengthen the cliques \eqref{clique_semplice_generale}. 

\begin{theorem}
Given $(t,b,\beta,l) \in \{\mathcal T, \mathcal B, \mathcal B, \mathcal L\}$ such that \eqref{condition_clique_general} is satisfied, with $l$ minimum power level satisfying \eqref{condition_clique_general} and $b \neq \beta$, the following cliques are valid inequalities
\begin{equation}\label{clique_aggregati_1A}
     \sum_{\lambda \in \mathcal L: \, \lambda \geq l} z_{b \lambda} + x_{t \beta} \leq 1 \qquad  b \in \mathcal B, t \in \mathcal T, \beta \in \mathcal B \setminus\{b\}.
\end{equation}
\end{theorem}
\begin{proof}
If $x_{t \beta} = 1$, then for each $\lambda \geq l$ we have $z_{b \lambda}=0$ otherwise the SINR$_{t \beta}$ constraint \eqref{sinr_ineq_general} is violated.
If instead $z_{b \lambda} = 1$ for one $\lambda \geq l$, then $x_{t \beta}=0$ ($\beta \neq b$) since the SINR$_{t \beta}$ constraint \eqref{sinr_ineq_general} is violated. Hence, we cannot have simultaneously that $\displaystyle\sum_{\lambda \in \mathcal L: \, \lambda \geq l} z_{b \lambda} = 1$ and $x_{t \beta} = 1$.

Moreover, inequalities \eqref{clique_semplice_generale} are implied by \eqref{clique_aggregati_1A} as
$$z_{bl} + x_{t \beta} \leq \sum_{\lambda \in \mathcal L: \, \lambda \geq l} z_{b \lambda} + x_{t \beta} \leq 1.$$
\end{proof} 
\begin{flushright}
$\square$
\end{flushright}

\begin{theorem}
Given $(t,b,\beta,l) \in \{\mathcal T, \mathcal B, \mathcal B, \mathcal L\}$ such that \eqref{condition_clique_general} is satisfied for all $\beta \neq b$, the following cliques are valid inequalities
\begin{equation}\label{clique_aggregati_1B}
     z_{b l} + \sum_{\beta \in \mathcal B \setminus \{b\}} x_{t \beta} \leq 1 \qquad  b \in \mathcal B, t \in \mathcal T, l \in \mathcal L.
\end{equation}
\end{theorem}
\begin{proof}

If $z_{b l} = 1$, then each $x_{t \beta}=0$ ($\beta  \neq b$) since the SINR$_{t \beta}$ constraint \eqref{sinr_ineq_general} is violated. If instead $x_{t \beta} = 1$ for one $\beta \neq b$, then $z_{b l}=0$ otherwise the SINR$_{t \beta}$ constraint \eqref{sinr_ineq_general} is violated.
 Hence, we cannot have simultaneously that $z_{b l} = 1$ and $\displaystyle\sum_{\beta \in \mathcal B\setminus \{b\}} x_{t \beta} = 1$.

Moreover, inequalities \eqref{clique_semplice_generale} are implied by \eqref{clique_aggregati_1B} as
$$z_{bl} + x_{t \beta} \leq z_{b l} + \sum_{\beta \in \mathcal B \setminus \{b\}} x_{t \beta} \leq 1.$$
\end{proof} 
\begin{flushright}
$\square$
\end{flushright}

\bigskip

Moreover, given the testpoint $t \in \mathcal T$, the server $b \in \mathcal B$ with power $P_l \in \mathcal P$ and the interferer $\beta \in \mathcal B$ such that the SINR measured in $t$ and that considers as the only interferer $\beta$ is always below the threshold $\delta$ for each possible $\beta \neq b$, i.e. 
\begin{equation*}
\frac{\tilde a_{t b}P_{l}}{\mu + \tilde a_{t \beta}P_j} < \delta \quad \forall \beta \in \mathcal B \setminus \{b\}, \forall P_j \in \mathcal P
\end{equation*}
which can be easily verified directly with
\begin{equation}
\label{condition_clique_2_general}
    \frac{\tilde a_{t b}P_{l}}{\mu + \tilde a_{th}P_{min}} < \delta, \,\quad \tilde a_{th} \coloneqq \min_{\beta \in \mathcal B \setminus \{b\}} \{\tilde a_{t \beta}\}
\end{equation}
we can exclude the possibility that $b$ is activated at a power level equal to $l$ and simultaneously $b$ serves $t$ using
\begin{equation}
\label{clique_semplice_2_generale}
    z_{bl} + x_{t b} \leq 1.
\end{equation}
Using \eqref{constr_power}, we can strengthen the cliques \eqref{clique_semplice_2_generale}.

\begin{theorem}
Given $(t,b,l) \in \{\mathcal T, \mathcal B, \mathcal L\}$ such that \eqref{condition_clique_2_general} is satisfied and $l$ corresponds to the
maximum power level such that \eqref{condition_clique_2_general} is satisfied, the following cliques are valid inequalities
\begin{equation}
     \sum_{\lambda \in \mathcal L: \, \lambda \leq l} z_{b \lambda} + x_{t b} \leq 1 \qquad  b \in \mathcal B, t \in \mathcal T.
     \label{clique_aggregati_2}
\end{equation}
\end{theorem}
\begin{proof}
If $x_{tb} = 1$, then for each $\lambda \leq l$  
we have $z_{b \lambda}=0$, otherwise the SINR$_{tb}$ constraint \eqref{sinr_ineq_general} is violated.
If instead $z_{b \lambda} = 1$ for one $\lambda \leq l$, then $x_{tb}=0$ since the SINR$_{tb}$ constraint \eqref{sinr_ineq_general} is violated.
Hence, we cannot have simultaneously that $\displaystyle\sum_{\lambda \in \mathcal L:\, \lambda \leq l} z_{b \lambda} = 1$ and $x_{t b} = 1$.

Moreover, inequalities \eqref{clique_semplice_2_generale} are implied by \eqref{clique_aggregati_2} as
    $$z_{bl} + x_{t b} \leq \sum_{\lambda \in \mathcal L: \, \lambda \leq l} z_{b \lambda} + x_{t b} \leq 1.$$
\end{proof}
\begin{flushright}
$\square$
\end{flushright}

\subsubsection{Variable upper bounds}
Variable upper bound constraints (VUBs) 
\begin{equation}
x_{tb} \leq \sum_{l \in  \mathcal{L}} z_{bl} \qquad t \in \mathcal T, b \in \mathcal B \label{VUBs}
\end{equation}
enforce that a testpoint $t \in \mathcal T$ can be assigned to a transmitter $b \in \mathcal B$ only if $b$ is activated, i.e., only if $b$ is using one strictly positive power level. They are known to strengthen the quality of linear relaxation significantly.

Let us denote by $\mathcal L_{tb} \subseteq \mathcal L$ the subset of power levels $l$ that satisfy \eqref{condition_clique_2_general} for a given transmitter-receiver pair $(b,t) \in \{\mathcal B,\mathcal T\}$, meaning that we can exclude that $b$ is activated at a power level $l \in \mathcal L_{tb}$ and simultaneously $b$ serves $t$. 
The VUBs \eqref{VUBs} can be tightened to 
\begin{equation}
    \label{tightened_vub}
    x_{tb} \leq \sum_{l \in \mathcal L \setminus \mathcal L_{tb}} z_{bl}  \qquad t \in \mathcal T, b \in \mathcal B 
\end{equation}

\subsection{Tightening procedure for the big-$M$}\label{ch_rcf}
To further reduce the model size, we propose a reduced cost fixing method, in short RCF (see \cite{achterberg2020presolve} for a survey on presolve techniques). Although this procedure is well-known and widespread, no one has ever tried to see its effects on this type of problem (based on our knowledge).

By solving the linear relaxation of the problem, we can get the lower bound $lb$  and the corresponding reduced costs $ \bar{c}_{bl}$ associated with the sole variables $z_{bl}$ in the optimal solution of the linear relaxation. Then, given an upper bound $ub> lb$, if 
$ \bar{c}_{ b l} \geq ub-lb\text{  for some }  b \in \mathcal B,  l \in \mathcal L$, the corresponding $z_{ b l}$ must be at its lower bound in every optimal solution; hence we can fix $z_{ b  l}=0$.
 
Whenever the fixing of a variable $z_{ b l}$ occurs at a given $ l\in \mathcal L$ such that $P_{ l} = P_{max}$, we can recompute and reduce the big-$M$, resulting in a tightening of the formulation. 
Indeed, after the RCF, we may have some transmitters $b$ that cannot emit at the maximum power level $l$ such that $P_l = P_{max}$, since the corresponding $z_{bl}$ variables have been fixed to zero. In such cases, the value by which $a_{tb}$ is weighted in the big-$M$ (see \eqref{bigM}) can be reduced to the highest power value that $b$ can assume, which is strictly less than $P_{max}$.

To formalize it, let us define the set $\mathcal B^R \subseteq \mathcal B$ of base stations affected by RCF, i.e., such that $b \in \mathcal B^R$ if the variable $z_{bl}$ has been fixed to zero for at least one $l \in \mathcal L$. Then, for a given $b \in \mathcal B^R$, we can define the set $\mathcal L_b^R \subset \mathcal L$ of power levels that $b$ can assume after the RCF. We denote by $P_{b, max}^R$ the power value corresponding to the maximum power level that $b \in \mathcal B^R$ can assume. Since $\mathcal L_b^R \subset \mathcal L$, we have that $P_{b, max}^R \leq P_{max}$.
Using this notation, we can write down the new value of the big-$M$
$$M^{\prime}_{t \beta} = \delta \mu + \delta \left(P_{max} \sum_{b \in  \mathcal{B} \setminus \{\beta, \mathcal{B}^R\}} \tilde a_{tb} + \sum_{b \in \mathcal{B}^R} P_{b, max}^R \tilde a_{tb}\right) = \delta \mu + \delta \sum_{b \in \mathcal{B}\setminus \{\beta\}} \tilde{P_b} \tilde a_{tb}$$
which satisfies 
$M_{t \beta} \geq M^{\prime}_{t \beta}$ since $P_{max} \geq \tilde P_b = \begin{cases} P_{max}  & \text{ if } b \in \mathcal B \setminus \mathcal B^R\\
P^{R}_{b,max} & \text{ if } b \in \mathcal B^R. \end{cases}$

The smaller the optimality gap given by the estimated lower and upper bounds, the greater the number of $z_{bl}$ variables that can be fixed to zero, and the smaller the big-$M$ coefficients.
Hence, applying a standard algorithm -- as implemented in MIP commercial solvers -- to the tightened formulation (i.e., the formulation got after the RCF) produces stronger bounds and a faster solution.

We can apply a further reduction. Since we have information on the maximum number of transmitters that can be installed from the $ub$, we can further reduce the value of the big-$M$ by replacing the sum of all the interfering signals in the testpoint $t$ with the sum of the \emph{strongest} interfering signals in $t$. In particular, only the strongest $\gamma$ interferers are considered, where $\gamma$ is the maximum number of transmitters that can be activated.
Given $\gamma$, the big-$M$ can be computed as
\begin{equation}
    \label{bigM_variable_final}
    M^{\prime\prime}_{t \beta} = \delta \mu + \delta \displaystyle \sum_{b \in  \mathcal{A}_t \setminus \{\beta\}} \tilde{P}_{b} \tilde a_{tb} \leq M^{\prime}_{t \beta}\leq M_{t \beta}
\end{equation}
where $\mathcal A_t \subset \mathcal B$ is the set of the $\gamma$ base stations emitting the strongest signals received in $t$, i.e $|\mathcal{A}_t| = \gamma$. The smaller is $\gamma$, and the smaller is the big-$M$; therefore, the estimate of $\gamma$ should be as accurate as possible.

We observe that getting a good lower bound is straightforward since constraints \eqref{VUBs} naturally lead to a good linear relaxation value. Conversely, finding a good upper bound is a more daunting task. Although commercial MIP solvers can be used to derive a feasible solution, it may be time-consuming.
Consequently, we derived a fixing heuristic based on the observation that the fractional values of the variables in the LP relaxation are often good predictors of zero/non-zero variables in an optimal ILP solution. This especially occurs when the LP relaxation is extremely tight.
Our fixing heuristic is made up of the following steps:
\begin{enumerate}
        \item Solve the LP relaxation of the problem and take the fractional solution
        \item Identify the set of fractional variables that are likely to be zero (i.e., whose value is less than a very low threshold in the fractional solution)
        \item Fix the selected variables to zero and perform bound strengthening  to propagate implications
    \item Solve the resulting ILP problem to get a near-optimal feasible solution.
    \end{enumerate}
We note that the variables rounded to zero correspond to the power levels that are not needed by the transmitters to cover the target area. 

\subsection{The final formulation}\label{ch_final_formulation_variable}
The final formulation (F) we propose differs from the initial formulation \eqref{eq:general} since:
\begin{enumerate}
    \item the number of servers and the number of levels for the power has been reduced according to the three operations described in Section \ref{ch_presolve_variable};
    \item it includes the addition of the VUBs \eqref{tightened_vub} and the cliques \eqref{clique_aggregati_1A}, \eqref{clique_aggregati_1B}, \eqref{clique_aggregati_2};
    \item it is the result of a reduced cost fixing procedure;
    \item the big-$M$ appearing in the SINR is formulated as in \eqref{bigM_variable_final}.
\end{enumerate}

\section{Computational Experience}
\label{ch_results}
The code has been implemented in Python and the experiments have been carried out on a Ubuntu server with an Intel(R) Xeon(R) Gold 5218 CPU running at 2.30 GHz, with 96 GB of RAM and 8 cores. Gurobi Optimizer 10.0.1 \cite{GurobiOptimizer} with default settings has been employed as an MIP solver. We set a time limit of four hours for computation time.

\subsubsection{Testbed}
The testbed is made of instances obtained from an existing 4G LTE 800 MHz network in the Municipality of Bologna (Italy) provided by the FUB. 
The system noise and the power received in each receiver by each possible transmitter have been simulated by the FUB using the \textit{Cost Hata} model \cite{akhpashev2016cost, gadze2020improved, rumney2013lte}.
The power values are in $W$ and scaled by a factor of $10^{10}$ to avoid numerical issues and obtain better accuracy on optimal solutions, as suggested in \cite{d2016towards}. Accordingly, the threshold on the quality of service $\delta$ and the system noise $\mu$ are expressed in $W$;  $\mu$ is also scaled by $10^{10}$. The emitted power values considered in this study are three: $20W$, $40W$, $80W$.
The threshold $\varepsilon$ has been set around -110 dBmW \cite{telco, usat}. In \eqref{constr_coverage}, the parameter $r_t$  has been set to the fraction of the population living in the elementary area $t \in \mathcal T$, and $r$ to the minimum fraction of the population to be covered by service. We used $|S_t| = 10$, namely we selected the strongest ten signals as possible servers for each testpoint $t$.

From this network, we derived 16 instances (BO1 to BO16), each of them having $|\mathcal B|= 135$ and $|\mathcal T| = 4\,693$. 
Each instance differs in the fraction of the population to be served ($r$) and the quality of service required (given by $\delta$, chosen in the typical range of LTE services), as reported in Table \ref{tab:instances_variable}. 
\begin{table} [h!] 
\caption[Characteristics of the instances of the variable-power case]{Characteristics of the instances: SINR threshold ($\delta$) and fraction of population to be served ($r$).} 
\begin{threeparttable}
\resizebox{\textwidth}{!}{
\begin{tabular}{lrrrrrrrrrrrrrrrr}
\textbf{Instance} & \textbf{BO1}   & \textbf{BO2}    & \textbf{BO3}     & \textbf{BO4}   & \textbf{BO5}  & \textbf{BO6}  & \textbf{BO7}  & \textbf{BO8}  & \textbf{BO9}  & \textbf{BO10} & \textbf{BO11} & \textbf{BO12}  & \textbf{BO13}  & \textbf{BO14} & \textbf{BO15} & \textbf{BO16}\\
\hline
$\delta${[}dBW{]} & -10 & -9,5 & -9 & -8.5  & -8 & -7.5 & -7  & -6.5 & -6 & - 5.5  & - 5 & -3 & 0 & +3 & +5 & +7 \\
r  & 1 & 1  & 1   & 1 & 1 & 1 & 1 & 1 & 1 & 0.99 & 0.99 & 0.95 & 0.85 & 0.75 & 0.70 & 0.65\\
\hline
\end{tabular}}
\end{threeparttable}
\label{tab:instances_variable}
\end{table}
The estimate of the upper bound has then been obtained on each instance using the fixing heuristic.

\subsubsection{Results}

In this section, we show the impact of the operations discussed in Section \ref{ch_contribution}. To this end, we compare the results obtained using formulation \eqref{eq:general} -- denoted as the basic formulation B -- and the final setting described in Section \ref{ch_final_formulation_variable}. Table \ref{tab:formulations_variable} provides an explicit description of the tested formulations.  We use three evaluation criteria, namely the size, the sparsity, and the quality of the gaps at the root node and the end of the optimization. 
\begin{table} [h!] 
\caption[Characteristics of the tested formulations for the variable-power case]{Characteristics of the tested formulations.}
\begin{threeparttable}
\resizebox{\textwidth}{!}{
\begin{tabular}{ll}
   \textbf{Formulation} & \textbf{Characteristics}\\
   \hline
    B & Basic formulation \eqref{eq:general}\\
    B+CPs & Formulation \eqref{eq:general} plus the addition of cutting planes (i.e., VUBs \eqref{tightened_vub} and cliques \eqref{clique_aggregati_1A}, \eqref{clique_aggregati_1B}, \eqref{clique_aggregati_2})\\
    F(RCF) & Final setting reported in Section \ref{ch_final_formulation_variable}, including RCF steps 3 and 4\\
    F & Final setting reported in Section \ref{ch_final_formulation_variable}, excluding RCF steps 3 and 4\\
   \hline
\end{tabular}}
\end{threeparttable}
\label{tab:formulations_variable}
\end{table}
We observe that we differentiate between two distinct final settings in our analysis: F(RCF), where the RCF procedure is applied along with a corresponding reduction of the big-$M$ value, and F, where these procedures are not included. This distinction is necessary as RCF can only be applied to certain instances. Specifically, instances requiring low to medium service values (BO1 to BO11) can successfully be solved using RCF. For instances with high service demands (BO12 to BO16, i.e., $\delta > - 5$ dBW), RCF is not a viable procedure, as both Gurobi and our heuristic fail to find good feasible solutions to be used in the RCF within a reasonable time. Hence, we apply setting F(RCF) when possible, otherwise we apply setting F.

\begin{figure}[h!]
\centering
\includegraphics[width=.49\linewidth]{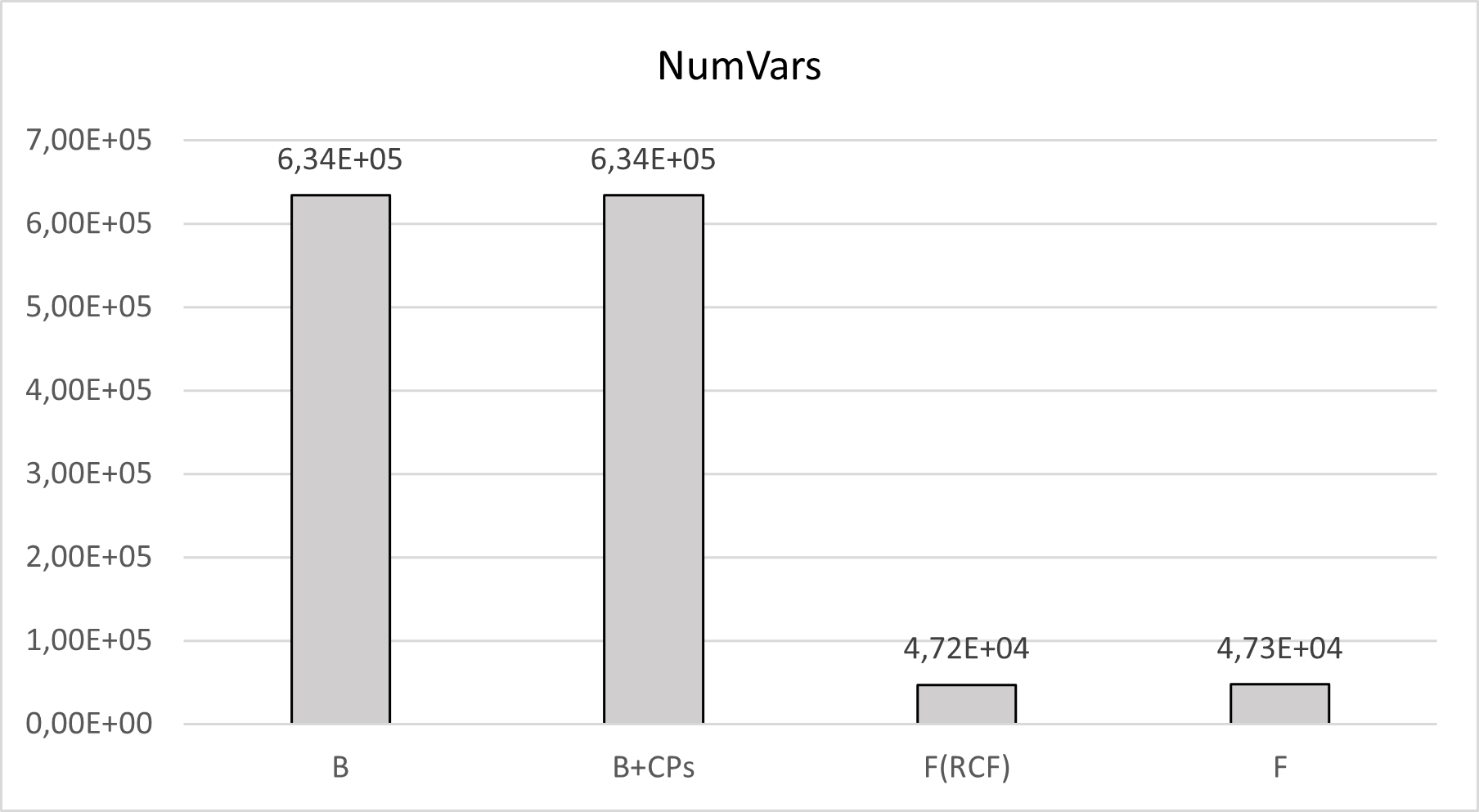} \includegraphics[width=.49\linewidth]{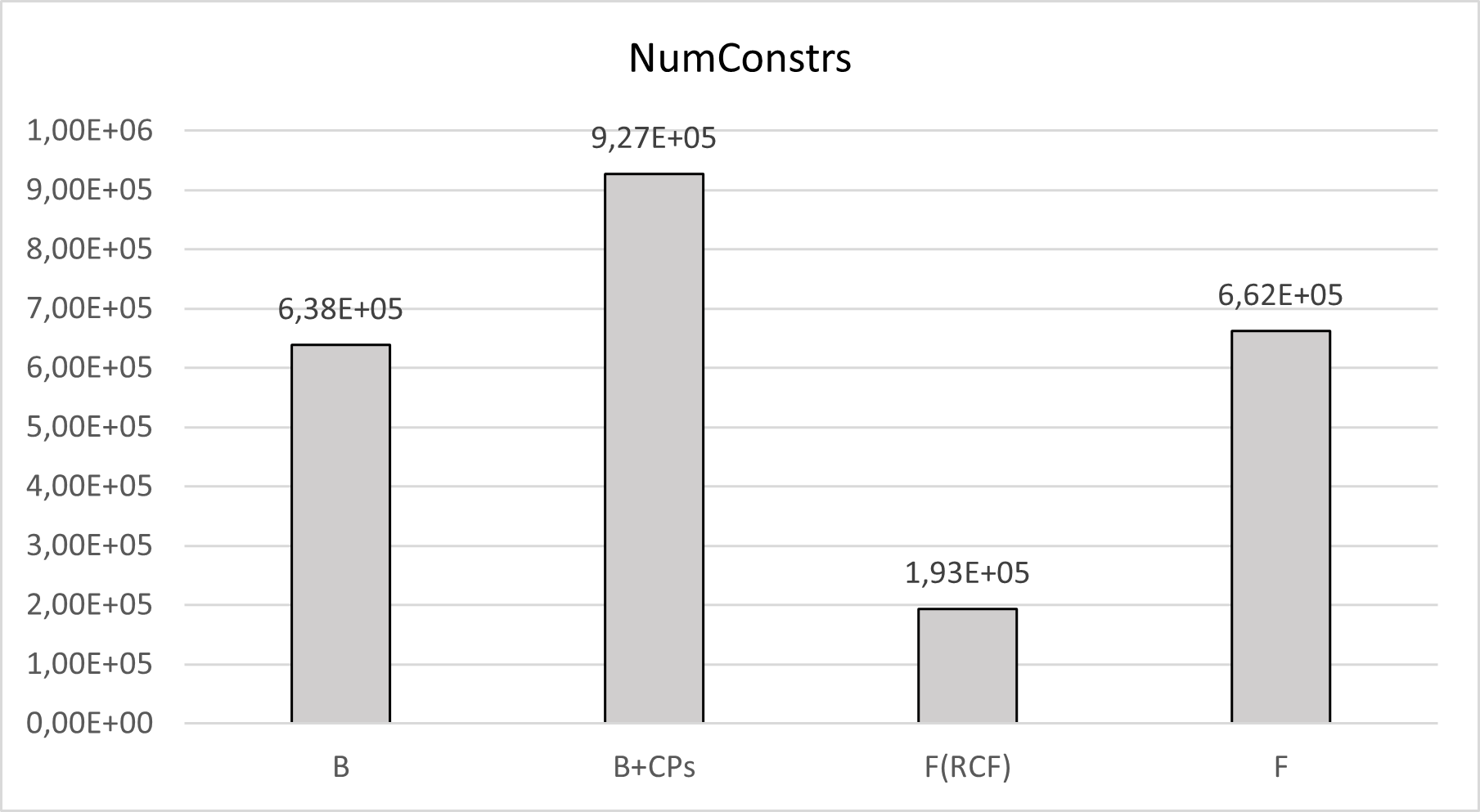}\\\includegraphics[width=.49\linewidth]{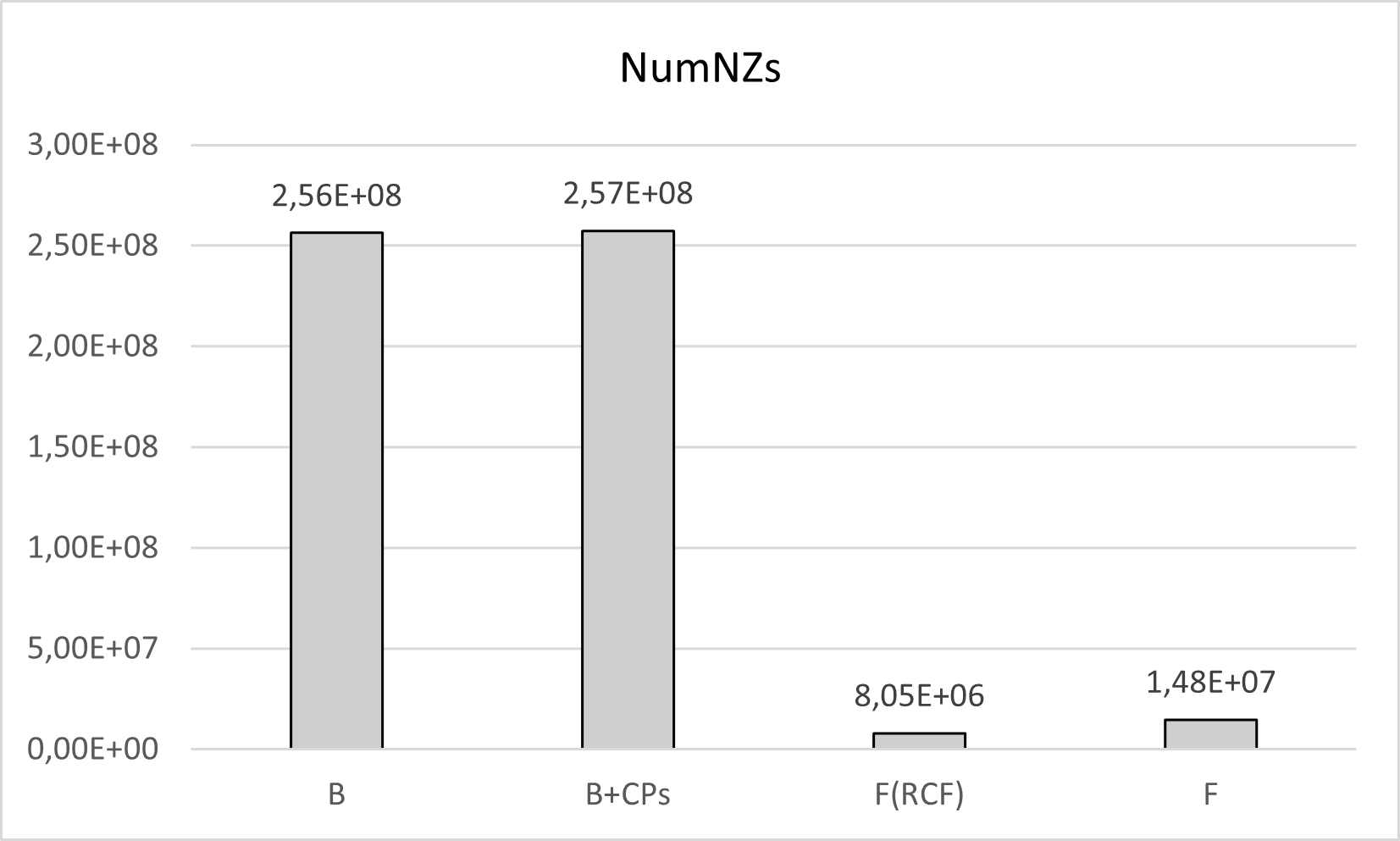}  
\caption[Average number of variables and constraints in the formulations of the variable-power case]{Average number of variables (top left), constraints (top right) and non-zeros (bottom) in the tested formulations.}
\label{fig:numVars_numCons_numNZs_var}
\end{figure}

\begin{figure}[H]
\centering
\includegraphics[width=0.49\linewidth]{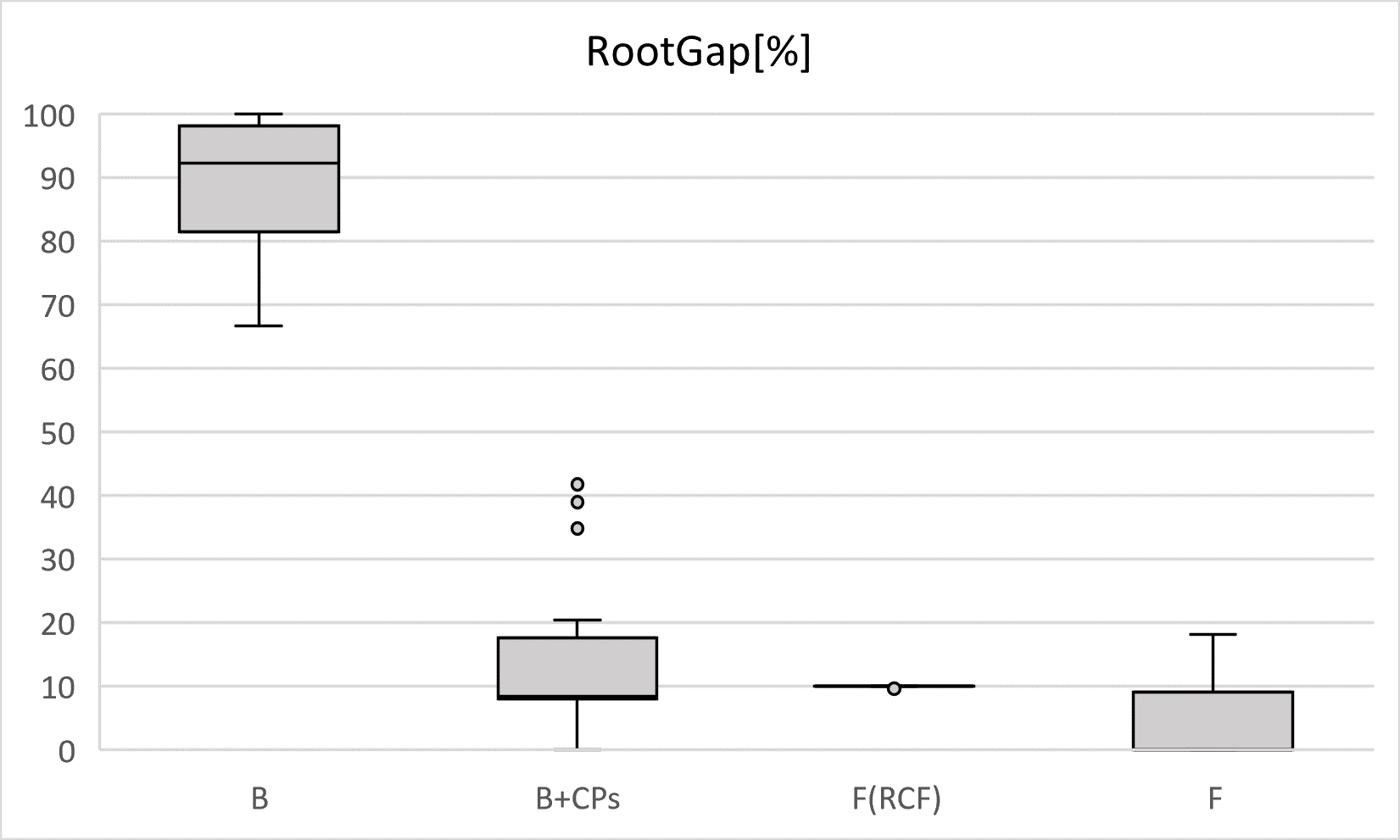}
\includegraphics[width=0.49\linewidth]{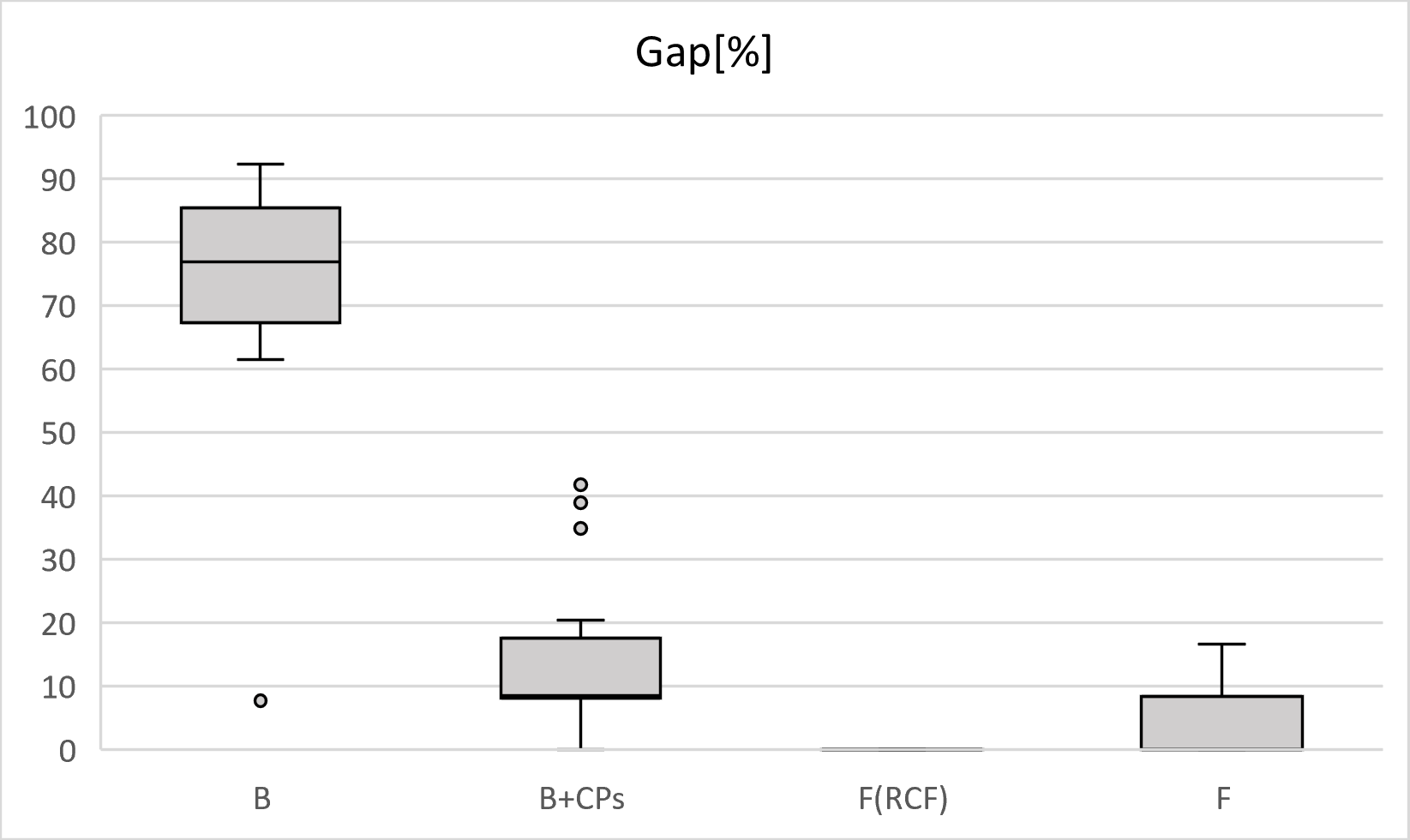}
\caption[Gap in the formulations of the variable-power case]{Box plots of the relative gap at the root node (on the left) and the end of the optimization phase (on the right) in the tested formulations (see Table \ref{tab:formulations_variable}).}
\label{fig:Gap_var}
\end{figure}

We report Figure \ref{fig:numVars_numCons_numNZs_var} showing the average number of variables, constraints and non-zeros of each formulation. Figure \ref{fig:Gap_var} instead show four box plots, one for each formulation, displaying the distribution of the data concerning the relative gap at the root node and the end of the optimization, based on a summary of five numbers (minimum value, first quartile, median value, third quartile, and maximum value).
From Figures \ref{fig:numVars_numCons_numNZs_var}-\ref{fig:Gap_var} we can claim that:
\begin{itemize}
    \item formulation B+CPs, leads to a significant improvement in the root relaxation value, affecting the gaps, revealing the decisive effect of adding the cutting planes;
    \item the aggressive reduction scheme in F(RCF), also including the RCF procedure, 
    leads to a definitely reduced and sparser formulation;
    \item the lighter reduction scheme in F, not including the RCF procedure, 
    leads to a reduced and sparser formulation compared to B+CPs, but slightly bigger and less sparse than F(RCF);
    \item the smallest gaps are achieved using F(RCF) or F; however, good final gaps can be reached already using B+CPs.
\end{itemize} 
The optimization results are reported in Table \ref{tab:B_vs_F_variable}. 
The evaluation metrics considered are: 
number of variables (NumVars), constraints (NumConstrs) and non-zeros (NumNZs); relative optimality gap at the root node (RootGap), final relative optimality gap (Gap); number of explored nodes (Nodes); total solution time (Time). 
Denoting the optimal value with Opt, the value of the lower bound at the root node with RootLB, and the value of the best lower bound with BestLB, the RootGap is computed as 100(Opt-RootLB)/Opt, and the Gap as 100(Opt-BestLB)/Opt. We reported in bold the gap of instances solved to optimality within the time limit.

Results show that formulations F(RCF) and F are definitely reduced in size and sparser than B. This is mainly due to the presolve operations and the reduced cost fixing procedure (when viable). The quality of the bounds at the root node is better in F(RCF) and F, mainly due to the addition of cutting planes. 
Thanks to a good root bound, the number of the explored nodes is heavily reduced in F(RCF) and F, and consequently, the time spent on the branch-and-cut tree search. Overall, solution times are significantly reduced if we compare F(RCF) or F to B. Indeed, none of the basic formulations have been successfully solved within the time limit of 4 hours, whereas all final formulation but one have been solved to optimality in less than 45 minutes. 

\begin{table}[H]\caption[Optimization results for the variable-power case]{Optimization results for the instances BO1--BO16. Time is expressed in seconds.}
\centering
\resizebox{0.8\textwidth}{!}{
\begin{tabular}{llrrrrrrr}
\textbf{Instance} & \textbf{Formulation} & \textbf{NumVars} & \textbf{NumConstrs} & \textbf{NumNZs} & \textbf{RootGap{[}\%{]}} & \textbf{Gap{[}\%{]}} & \textbf{Nodes} & \textbf{Time{[}s{]}} \\ \hline
BO1               & \textbf{B}           & 633960           & 638383              & 2,56E+08        & 100,00                   & 92,31                & 1484           & TL                   \\
                  & \textbf{B+CPs}       & 633960           & 750990              & 2,57E+08        & 7,68                     & \textbf{0,00}        & 74             & 14354,51             \\
                  & \textbf{F(RCF)}      & 47201            & 164365              & 8,17E+06        & 9,97                     & \textbf{0,00}        & 83             & 429,96               \\ \hline
BO2               & \textbf{B}           & 633960           & 638383              & 2,56E+08        & 100,00                   & 7,69                 & 1503           & TL                   \\
                  & \textbf{B+CPs}       & 633960           & 755652              & 2,57E+08        & 8,06                     & 8,06                 & 1              & TL                   \\
                  & \textbf{F(RCF)}      & 47201            & 169027              & 8,19E+06        & 9,97                     & \textbf{0,00}        & 110            & 534,67               \\ \hline
BO3               & \textbf{B}           & 633960           & 638383              & 2,56E+08        & 100,00                   & 69,23                & 14472          & TL                   \\
                  & \textbf{B+CPs}       & 633960           & 760640              & 2,57E+08        & 8,32                     & 8,32                 & 1              & TL                   \\
                  & \textbf{F(RCF)}      & 47201            & 174015              & 8,20E+06        & 9,97                     & \textbf{0,00}        & 122            & 518,33               \\ \hline
BO4               & \textbf{B}           & 633960           & 638383              & 2,56E+08        & 92,31                    & 76,92                & 14505          & TL                   \\
                  & \textbf{B+CPs}       & 633960           & 765973              & 2,57E+08        & 8,17                     & 8,17                 & 1              & TL                   \\
                  & \textbf{F(RCF)}      & 47201            & 179348              & 8,22E+06        & 9,97                     & \textbf{0,00}        & 129            & 471,71               \\ \hline
BO5               & \textbf{B}           & 633960           & 638383              & 2,56E+08        & 100,00                   & 76,92                & 12182          & TL                   \\
                  & \textbf{B+CPs}       & 633960           & 771677              & 2,57E+08        & 8,32                     & 8,32                 & 1              & TL                   \\
                  & \textbf{F(RCF)}      & 47201            & 185052              & 8,27E+06        & 9,97                     & \textbf{0,00}        & 105            & 541,17               \\ \hline
BO6               & \textbf{B}           & 633960           & 638383              & 2,56E+08        & 92,31                    & 76,92                & 15032          & TL                   \\
                  & \textbf{B+CPs}       & 633960           & 777637              & 2,57E+08        & 8,45                     & 8,45                 & 1              & TL                   \\
                  & \textbf{F(RCF)}      & 47200            & 190259              & 8,24E+06        & 9,97                     & \textbf{0,00}        & 151            & 680,76               \\ \hline
BO7               & \textbf{B}           & 633960           & 638383              & 2,56E+08        & 92,31                    & 61,54                & 14708          & TL                   \\
                  & \textbf{B+CPs}       & 633960           & 784138              & 2,57E+08        & 8,49                     & 8,49                 & 1              & TL                   \\
                  & \textbf{F(RCF)}      & 47200            & 197513              & 8,27E+06        & 9,97                     & \textbf{0,00}        & 293            & 586,75               \\ \hline
BO8               & \textbf{B}           & 633960           & 638383              & 2,56E+08        & 92,31                    & 84,62                & 14353          & TL                   \\
                  & \textbf{B+CPs}       & 633960           & 791167              & 2,57E+08        & 8,41                     & 8,41                 & 1              & TL                   \\
                  & \textbf{F(RCF)}      & 47201            & 204542              & 8,32E+06        & 9,97                     & \textbf{0,00}        & 127            & 598,99               \\ \hline
BO9               & \textbf{B}           & 633960           & 638383              & 2,56E+08        & 92,31                    & 69,23                & 14793          & TL                   \\
                  & \textbf{B+CPs}       & 633960           & 798671              & 2,57E+08        & 8,53                     & 8,53                 & 1              & TL                   \\
                  & \textbf{F(RCF)}      & 47202            & 212046              & 8,39E+06        & 9,97                     & \textbf{0,00}        & 86             & 537,43               \\ \hline
BO10              & \textbf{B}           & 633960           & 638384              & 2,57E+08        & 90,00                    & 90,00                & 14756          & TL                   \\
                  & \textbf{B+CPs}       & 633960           & 806851              & 2,57E+08        & 9,01                     & 9,01                 & 1              & TL                   \\
                  & \textbf{F(RCF)}      & 47158            & 220226              & 7,11E+06        & 9,63                     & \textbf{0,00}        & 37             & 1922,65              \\ \hline
BO11              & \textbf{B}           & 633960           & 638384              & 2,57E+08        & 90,00                    & 90,00                & 14753          & TL                   \\
                  & \textbf{B+CPs}       & 633960           & 815638              & 2,57E+08        & 4,54                     & 4,54                 & 1              & TL                   \\
                  & \textbf{F(RCF)}      & 47157            & 229013              & 7,16E+06        & 9,63                     & \textbf{0,00}        & 37             & 2311,70              \\ \hline
BO12              & \textbf{B}           & 633960           & 638384              & 2,57E+08        & 85,71                    & 85,71                & 24276          & TL                   \\
                  & \textbf{B+CPs}       & 633960           & 860727              & 2,57E+08        & 0,00                     & \textbf{0,00}        & 1              & 10587,45             \\
                  & \textbf{F}           & 47335            & 274102              & 1,33E+07        & 0,00                     & \textbf{0,00}        & 1              & 181,99               \\ \hline
BO13              & \textbf{B}           & 633960           & 638384              & 2,57E+08        & 80,00                    & 80,00                & 14964          & TL                   \\
                  & \textbf{B+CPs}       & 633960           & 981892              & 2,58E+08        & 20,42                    & 20,42                & 1              & TL                   \\
                  & \textbf{F}           & 47335            & 395267              & 1,38E+07        & 18,20                    & 16,60                & 3659           & TL                   \\ \hline
BO14              & \textbf{B}           & 633960           & 638384              & 2,57E+08        & 75,00                    & 75,00                & 14641          & TL                   \\
                  & \textbf{B+CPs}       & 633960           & 1227228             & 2,59E+08        & 38,90                    & 38,90                & 1              & TL                   \\
                  & \textbf{F}           & 47335            & 640603              & 1,46E+07        & 0,00                     & \textbf{0,00}        & 1              & 2128,19              \\ \hline
BO15              & \textbf{B}           & 633960           & 638384              & 2,57E+08        & 66,67                    & 66,67                & 14800          & TL                   \\
                  & \textbf{B+CPs}       & 633960           & 1455551             & 2,60E+08        & 34,83                    & 34,83                & 1              & TL                   \\
                  & \textbf{F}           & 47327            & 867412              & 1,55E+07        & 0,00                     & \textbf{0,00}        & 1              & 1503,43              \\ \hline
BO16              & \textbf{B}           & 633960           & 638384              & 2,57E+08        & 66,67                    & 66,67                & 15068          & TL                   \\
                  & \textbf{B+CPs}       & 633960           & 1729581             & 2,61E+08        & 41,77                    & 41,77                & 1              & TL                   \\
                  & \textbf{F}           & 47281            & 1132474             & 1,67E+07        & 0,00                     & \textbf{0,00}        & 1              & 2556,58     \\ \hline         
\end{tabular}}
\begin{tablenotes}
\item TL, time limit reached
\end{tablenotes}
\label{tab:B_vs_F_variable}
\end{table}

Regarding the comparison between the final settings and B+CPs, we observe that the final setting leads to smaller and sparser formulations, and also to faster solution times. Indeed, with B+CPs we can solve only one instance to optimality within the time limit. 
\\In the end, setting F(RCF), or F in case RCF is not viable, turns out to be much more competitive than B or B+CPs.

In Tables \ref{tab:times_f_variable} we also report the computational time of every phase of the setting F(RCF). In particular, by LBTime we denote the time needed to get the lower bound given by the linear relaxation of the problem, by UBTime the time needed to get an upper bound using our fixing heuristic, by SolTime the time needed to solve the problem after the RCF procedure, and by Time the overall solution time given by the sum of the previous components. The efficiency in solving the instances using the scheme based on RCF is attributed to a rapid computation of the lower and upper bounds.
Both depend on the good quality of the linear relaxation of the problem, achieved thanks to the inclusion of the cuts we introduced. Note that for higher quality values, unfortunately the lower bound gets slightly worse and it is no longer possible to exploit our fixing heuristic. We also tried to exploit Gurobi as a heuristic to obtain an upper bound on high service instances, but unfortunately the bounds produced by Gurobi in times compatible with the use of this procedure were not good enough to actually apply the RCF.

\begin{table}[h!]\caption[Detailed solution times on the final formulation]{Detailed solution times on the instances BO1--BO11  where we can apply RCF. Time is expressed in seconds.}
\centering
\begin{tabular}{lrrrr}
\textbf{Instance} & \textbf{LBTime{[}s{]}} & \textbf{UBTime{[}s{]}} & \textbf{SolTime{[}s{]}} & \textbf{Time{[}s{]}} \\ \hline
BO1              & 34,85                  & 31,53                  & 363,58                  & 429,96               \\
BO2              & 33,61                  & 34,56                  & 466,50                  & 534,67               \\
BO3              & 32,79                  & 34,88                  & 450,66                  & 518,33               \\
BO4              & 37,57                  & 41,40                  & 392,74                  & 471,71               \\
BO5              & 33,25                  & 42,21                  & 465,71                  & 541,17               \\
BO6              & 39,08                  & 47,49                  & 594,19                  & 680,76               \\
BO7              & 34,14                  & 52,14                  & 500,47                  & 586,75               \\
BO8              & 33,75                  & 54,54                  & 510,70                  & 598,99               \\
BO9              & 36,43                  & 64,36                  & 436,64                  & 537,43               \\
BO10             & 47,49                  & 28,61                  & 1846,55                 & 1922,65              \\
BO11             & 52,87                  & 29,39                  & 2229,44                 & 2311,70\\ 
\hline
\end{tabular}
\label{tab:times_f_variable}
\end{table}

Finally, Table \ref{tab:bigM} compares the maximum values of the big-$M$ in B and F(RCF). We observe that the value of the largest big-$M$ used in F(RCF) is half that used in B in half the instances.

\begin{table}[H]\caption[Maximum big-$M$ values]{Maximum big-$M$ values on the instances BO1--BO11 where we can apply RCF.}
\centering
\resizebox{0.32\textwidth}{!}{\begin{tabular}{llr}
\textbf{Instance} & \textbf{Formulation} & \multicolumn{1}{r}{\textbf{MaxBigM}} \\ \hline
BO1               & \textbf{B}                    & 91958,73                             \\
                  & \textbf{F(RCF)}               & 45979,28                             \\\hline
BO2               & \textbf{B}                    & 103179,39                            \\
                  & \textbf{F(RCF)}               & 103179,20                            \\\hline
BO3               & \textbf{B}                    & 115769,18                            \\
                  & \textbf{F(RCF)}               & 115768,96                            \\\hline
BO4               & \textbf{B}                    & 129895,16                            \\
                  & \textbf{F(RCF)}               & 64947,46                             \\\hline
BO5               & \textbf{B}                    & 145744,77                            \\
                  & \textbf{F(RCF)}               & 145744,49                            \\\hline
BO6               & \textbf{B}                    & 161284,63                            \\
                  & \textbf{F(RCF)}              & 80642,16                             \\\hline
BO7               & \textbf{B}                    & 183481,79                            \\
                  & \textbf{F(RCF)}               & 183481,44                            \\\hline
BO8               & \textbf{B}                   & 205869,96                            \\
                  & \textbf{F(RCF)}               & 205869,56                            \\\hline
BO9               & \textbf{B}                    & 230989,89                            \\
                  & \textbf{F(RCF)}               & 115494,79                            \\\hline
BO10              & \textbf{B}                    & 259174,92                            \\
                  & \textbf{F(RCF)}               & 259174,08                            \\\hline
BO11              & \textbf{B}                    & 290799,04                            \\
                  & \textbf{F(RCF)}               & 142654,51   \\\hline                        
\end{tabular}}
\label{tab:bigM}
\end{table}

\section{Conclusions}
\label{ch_conclusions}


In this paper we worked on improving the natural formulation proposed in the literature of wireless network design for the site and power assignment problem. Specifically, we provided several presolve operations to reduce the number of problem variables and overall problem size, along with valid cliques and variable upper bounds to reduce solution times. We also proposed an aggressive reduction scheme based on a reduced cost fixing procedure that reduces the big-M values, strengthening the formulation and reducing the problem size. 
Computational tests were conducted on authentic 4G LTE data from the Municipality of Bologna in Italy. The results confirmed the efficacy of our proposals, which, compared to the standard solution of traditional problem formulation, resulted in faster solution times. Indeed, only with our proposals we could efficiently solve large instances of the problem to optimality, in solution times consistent with planning operations. 
However, the reduced cost fixing procedure can be used only when proper upper bounds can be computed quickly: this is not trivial in all cases, particularly when high-quality service is required.

\section*{Acknowledgments}
The authors gratefully acknowledge the support of Fondazione Ugo Bordoni, in particular Manuel Faccioli, Federica Mangiatordi, Luca Rea, Guido Riva, Pierpaolo Salvo, and Antonio Sassano.

%
%


\end{document}